\documentstyle[preprint,aps,eqsecnum]{revtex}
\begin{document}
\draft
\title{Coherent states, Path integral, and Semiclassical
approximation}
\author{Kunio FUNAHASHI, Taro KASHIWA, and Seiji SAKODA}
\address{Department of Physics, Kyushu University,Fukuoka 812, Japan}
\author{Kazuyuki  FUJII}
\address{Department of Mathematics, Yokohama
 City University, Yokohama 236, Japan}
\date{\today}
\maketitle
\begin{abstract}
Using the generalized coherent states we argue that the path integral
formulae for $SU(2)$ and $SU(1,1)$ (in the discrete series) are WKB
exact,
if the starting point is expressed as the trace of $e^{-iT\hat H}$
with $\hat
H$ being given by a linear combination of generators. In our case,
WKB
approximation is achieved by taking a large ``spin'' limit:
$J,K\rightarrow
\infty$. The result is obtained directly by knowing that the each
coefficient
vanishes under the $J^{-1}$($K^{-1}$) expansion
and is examined by another method to be legitimated.
We also point out that the discretized form of path integral is
indispensable,
in other words, the continuum path integral expression leads us to a
wrong
result. Therefore a great care must be taken when some geometrical
action
would be adopted, even if it is so beautiful, as the starting
ingredient of
path integral.
\end{abstract}


\section{Introduction}
Physical systems in actual situation are so complicated that even in
a
simplified form there usually need some approximation techniques,
such as
perturbation of the coupling constant or the self-consistent
manner of Hartree-Fock. In path integral the most promising one seems
to
be a semiclassical (WKB) approximation. However there are some
systems
in which the WKB approximation gives the exact result:
a harmonic oscillator is the well-known example (and may be the only
case if we are in a usual quantum mechanical circumstance, that is,
on a flat
and non-compact manifold); since the Hamiltonian is quadratic in
 momenta and coordinates, yielding a Gaussian form in the path
integral.

If generalization is made to quantum mechanics on a non-trivial
manifold
such as $SU(2)$ spin system\cite{NR}\cite{TK}\cite{KJ} or
${\bf C}P^N$ system,  a new possibility may occur \cite{SN}: this new
possibility of exactness of the WKB approximation is discussed in
connection
with the theorem of Duistermaat-Heckman(D-H) \cite{DH} and is
extended to
Grassmannian manifold \cite{RR}. So far the discussion has been
concentrated
mainly on the geometrical view point \cite{AF}.
However there are a number of unsatisfactory points and mysteries:
the
ingredient for obtaining path integral formulae  is (apart from
\cite{NR},\cite{TK}, and
\cite{KJ})  so called (generalized) coherent states. They have
employed naive
calculus,
$
\langle g | g' \rangle \sim 1 + \langle  g | \delta g \rangle \sim
\exp \langle  g | \delta g  \rangle ,
$
where $|g \rangle $ is some generalized coherent states and the
elements
$g$ and $g'$ are assumed to $g' = g + \delta g$ with $\delta g \ll
1$; which,
however, cannot be justified because $g$ and $g'$ are the integration
variables in the path integral. Thus obtained ``quantum" action in
the path
integral formula has already been in a semiclassical state. The
reason why
this kind of rough estimation could have been accepted is that the
resultant
action is beautifully geometric. In this sense, it is still
unsatisfactory for
us that ad hoc adoption of geometrical actions in path integral even
if those
would give us a correct result.

Another issue in our mind is that we should be so careful in
performing WKB approximation: first point is so called
overspecification problem. It is often said that the WKB
approximation
is not allowed in the case of kernel under the (canonical) coherent
state representation;
\begin{eqnarray}
a| z \rangle & = & z | z\rangle;   \quad
|z \rangle \equiv \exp \left( -{1\over 2} |z|^{2}+za^{\dagger}
\right)|0
\rangle, \quad ( a|0\rangle=0 )\  ,\nonumber \\
\langle z|z^{\prime} \rangle &=&\exp\left({-{1\over 2}{\vert
z\vert}^{2}-{1\over 2}{\vert
z^{\prime}\vert}^{2}+z^{*}z^{\prime}}\right)\ ,
\label{intrb}
\end{eqnarray}
\begin{equation}
\int{{dz^{*}dz}\over\pi}\left\vert{z}\right\rangle
\left\langle{z}\right\vert={\bf 1} ,\label{intrc}
\end{equation}
giving
\begin{eqnarray}
& &K(z_F,z_I;T) \equiv \langle z_F \vert e^{-iTH(a^\dagger,a)}
\vert z_I \rangle =\lim_{N \rightarrow \infty} \langle z_F \vert
\left({\bf 1}  -i\epsilon H(a^\dagger,a)\right)^N \vert z_I
\rangle \quad \left(\epsilon
\equiv {T\over N}\right) \label{intrd} \\
&=& \lim_{ N \rightarrow \infty}
\prod_{j=1}^{N-1} \int {{dz_{j}^{*}dz_{j}}\over\pi}
\exp \left[ {- \sum_{j=1}^N  \left[{{1 \over 2}\left\{
z_{j}^{*}\left(z_{j}-z_{j-1}\right) -\left(z_{j}^{*}-z_{j-1}^{*}
\right) z_{j-1}\right\}
+ i\epsilon H(z_j^*, z_{j-1}) }\right]} \right]\nonumber\\
& &\mathop{\hphantom{\lim_{ N \rightarrow \infty}
\prod_{j=1}^{N-1} }}
\left(\hbox{with }{z_0=z_I ,\ z_N=z_F}\right)
\nonumber
\end{eqnarray}
where we have successively inserted the resolution of
unity (\ref{intrc}) into the product in third expression of the first
line.
The WKB-approximation is performed by putting $z_j=z_j^c+z_j^\prime$,
where
$z_j^c$'s satisfy the classical {\sl continuum} equations of motion,
$i \partial z^c\left( t \right) / \partial t = \partial H \left( z^*
, z
\right) / \partial {z^c\left( t \right)}^*$; since $z_j^c$'s are not
the
integration variables so that the continuum limit can be taken
safely.
However there are {\sl two} boundary conditions ; $z^c \left( T
\right)
= z_F ,\  z^c \left( 0 \right) = z_I$ despite the fact that the
equations of
motion are first order differential equations  \cite{SF} \footnote{On
the
other hand, when applied to the discrete time formulation given by
the second
line of (\ref{intrd}), WKB approximation is nothing but a saddle
point
approximation for $N-1$ integration variables. For this case we have
a set of
$2(N-1)$ equations
\[
i\left(z_{j}-z_{j-1}\right)={{\partial}\over{\partial
z_{j}^*}}H(z_{j}^*,\ z_{j-1})\ ,
-i\left(z_{j+1}^{*}-z_{j}^{*}\right)={{\partial}\over{\partial
z_{j}}}H(z_{j+1}^*,z_{j})\quad (1\le j\le N-1)\ .
\]
They should be solved with conditions $z_{0} = z_I$ for the first
equation and
$z_{N}^{*} = z_{F}^{*}$ for the second one respectively. Thus we see
there is
no such problem if we work with the discrete time formulation of path
integral
. }.
For the trace formula,
\begin{eqnarray}
{\rm Tr} e^{-iTH(a^\dagger, a) } &=& \int{{dz^{*}dz}\over\pi} \langle
z |
e^{-iTH(a^\dagger, a) } |z\rangle \label{intre}\\
&=& \lim_{ N \rightarrow \infty}
\prod_{j=1}^{N} \int \limits_{\rm PBC} \!
{{dz_{j}^{*}dz_{j}}\over\pi}
\exp\left\{  - z_{j}^{*}\left(z_{j}-z_{j-1}\right) -i \epsilon
H(z_{j}^{*},
z_{j-1})  \right\} ,\nonumber
\end{eqnarray}
the boundary condition becomes a periodic one; $z_{N} = z_{0}$(PBC),
which has of course no problem. Case is unchanged for the generalized
coherent state representation; however,  seems a little bit drastic
in $SU(2)$
of Nielsen-Rohrlich formula \cite{NR},\cite{TK},\cite{KJ}:
 the boundary condition is
$\phi(T) = \phi(0) + 2n\pi   \   ( n \ \in  \ {\bf Z} ),  $
but the equation of motion reads
$\dot \phi(t) =  h    \  (   h : {\rm constant} ). $
They are  never compatible. (See the discussion.)

The second point that we would like to mention is subtlety of the use
of
continuum path integral under the coherent representation: as an
example, take the harmonic oscillator, $H= \omega a^{\dagger}a$,  in
(\ref{intre}) and go to the naive continuum limit, giving
\begin{eqnarray}
{\rm Tr} e^{-i\omega a^{\dagger}a T }
 & = & \int\limits_{\rm PBC}\!  {{\cal D} z^{*}{\cal D} z}  \exp
\left\{  - \int_0^T  dt \left(   z^{*} \dot z + i\omega z^{*}z
\right)  \right\} \nonumber \\
 & \equiv &{\det \left( {d \over dt} + i
\omega \right)}^{-1} = {1\over{2i\sin(\omega T/2)}}
\quad,\label{intrf}
\end{eqnarray}
where the determinant has been defined as,
$\det ( d / dt  + i \omega ) \equiv  \prod_{n= -\infty}^{\infty}
( i 2n\pi / T  + i \omega ) $,
with $i2n\pi/T$ coming from a periodic eigenfunction;
$f_n(t) = e^{i2n\pi/T} / \sqrt T$,
and use has been made of the $\zeta$-function regularization. The
correct
answer is
\begin{equation}
{\rm Tr} e^{-i\omega a^{\dagger}a T } ={1\over{1-e^{-i\omega T}}}  =
{e^{i
\omega T/2} \over {2i\sin(\omega T/2)} }\ , \label{intrh}
\end{equation}
of course.
This discrepancy when being in the continuum expression is always
left in
other cases such as  $SU(2)$  and might be in the
Chern-Simons field theory as a need for the Coxeter number
\cite{EW}(see the discussion).

Motivated by these, we shall in this paper study the
exactness of the WKB approximation under the path integral formula in
the
case of $SU(2)$ and
$SU(1, 1)$ with the aid of generalized coherent states. To avoid the
above
questionable issues, we shall concentrate only on formulae like
(\ref{intre}).
In section 2, we shall introduce the path integral expression for the
character formulae of $SU(2)$ and $SU(1, 1)$ and study their
structure. The
following section 3 deals with the exactness of the WKB
approximation. The
final section will be devoted to  discussions and in the appendix,
geometrical
properties  of  generalized coherent states and the relationship to
the
canonical coherent states will be presented, which will be useful for
an analysis of ${\bf C}P^N$ as well as Grassmannian manifold.

\section{Coherent states and path integral formulae
\protect\\ for $SU(2)$ and $SU(1,1)$}

\subsection{$SU(2)$;}
$su(2)$ algebra reads
\begin{equation}
 [J_{+},J_{-}]=2J_3  ,\quad [J_3 , J_{\pm} ] = \pm J_{\pm} ,
\end{equation}
where $J_{\pm} \equiv J_1  \pm i J_2$. Take a representation as
usual,
\begin{equation}
 J_{3}\left\vert{J,\ M}\right\rangle= M \left\vert{J,\
M}\right\rangle , \  J_{\pm} \left\vert{J,\ M }\right\rangle = \sqrt{
(J
\mp M) ( J \pm M +1) } \left\vert{J,\ M \pm 1 }\right\rangle, \  (
|M|
\leq J )  .\label{sutwoalg}
\end{equation}
Let us define
\begin{equation}
\left\vert{\xi}\right) \equiv  e^{\xi J_{+}}\left\vert{J,\
-J}\right\rangle, \quad
\left\vert{\xi}\right\rangle \equiv {1\over{\left({\xi\vert
\xi}\right)^{1/2}}}\left\vert{\xi}\right)\quad,\quad\xi\in{\bf
C}\quad,\label{sutoa}
\end{equation}
where we have introduced $|J, \ -J \rangle$ as the fiducial vector
\cite{KS}.
Explicitly
\begin{equation}
\left\vert{\xi}\right)=  \sum_{m=0}^{2J}\xi^{m}{2J
\choose m}^{1/2}\left\vert{J, -J+m}\right\rangle, \label{sutob}
\end{equation}
and whose norm is found
to be
$
\left({\xi\vert \xi}\right)=\left({1+{\vert
\xi\vert}^{2}}\right)^{2J},\
$
giving the normalized state,
\begin{equation}
\left\vert{\xi}\right\rangle={1\over{\left({1+{\vert
\xi\vert}^{2}}\right)^{J}}}
\sum_{m=0}^{2J}\xi^{m}{2J\choose m}^{1/2}\left\vert{J,\
-J+m}\right\rangle\quad. \label{sutoc}
\end{equation}
They satisfy
\begin{equation}
\left\langle{{\xi}\left\vert{\xi^{\prime}}\right\rangle}\right.
= {{\left({1+\xi^{*}\xi^{\prime}}\right)^{2J}}\over{\left({1+{\vert
\xi\vert}^{2}}\right)^{J}\left({1+{\vert \xi^{\prime}\vert}^{2}}
\right)^{J}}}\ ,\label{sutoda}
\end{equation}
\begin{equation}
{{2J+1}\over\pi}\int{{d\xi^{*}d\xi}\over{\left(
1+|\xi|^2
\right)^{2}}}
\left\vert{\xi}\right\rangle\left\langle{\xi}\right\vert \equiv \int
d\mu(\xi^{*},\xi)
\left\vert{\xi}\right\rangle\left\langle{\xi}\right\vert
= {\bf 1}_{J}\quad, \label{sutodb}
\end{equation}
where
\begin{equation}
d\xi^* d\xi\equiv d{\rm Re}(\xi) d{\rm Im}(\xi) ,
\end{equation}
and
\begin{equation}
{\bf 1}_{J}\equiv\sum^J_{M=-J}\left\vert{J,\
M}\right\rangle\left\langle{J,\ M}\right\vert
\quad,
\end{equation}
is the identity operator in $2J+1$-dimensional irreducible
representation.  Matrix elements of generators are  found to be
\begin{eqnarray}
\left\langle {\left. \xi  \right|J_3\left| {\xi '}\right.}
\right\rangle &=&-J{{1-\xi ^*\xi '} \over {1+\xi ^*\xi '}}
\left\langle {\xi }
\mathrel{\left | {\vphantom {\xi  {\xi '}}} \right.
\kern-\nulldelimiterspace}
{{\xi '}} \right\rangle\quad, \hfill\nonumber\\
  \left\langle {\left. \xi  \right|J_+\left| {\xi '} \right.}
\right\rangle
&=&J{{2\xi ^*} \over {1+\xi ^*\xi '}}\left\langle {\xi }
\mathrel{\left | {\vphantom {\xi  {\xi '}}} \right.
\kern-\nulldelimiterspace} {{\xi '}}
\right\rangle\quad. \label{mele}
\end{eqnarray}

Armed with these machineries, we can now discuss the path integral
formula for a Hamiltonian
\begin{equation}
H=h_1J_1+h_2J_2+h_3J_3\in su(2)\ .
\end{equation}
The starting point is the trace formula
\begin{equation}
Z(T) \equiv {\rm Tr} e^{ -iHT} = {\rm Tr} \lim_{N \rightarrow \infty}
({\bf 1} - i\epsilon H)^N = \lim_{N \rightarrow \infty} Z_N,
\label{tracecha}
\end{equation}
where
\begin{equation}
Z_N \equiv  \prod_{j=1}^N \int \limits_{\rm PBC} \!
d\mu (\xi_j^*,\xi_j) \exp \left[ i \left\{ -i \ln \left\langle \xi_j
\left\vert
\xi_{j-1} \right\rangle \right.
-\epsilon {\cal H} \left( \xi_j^*,\xi_{j-1} \right) \right\}\right]
\quad,\label{sdisca}
\end{equation}
with $d\mu(\xi_{j}^{*},\xi_{j}) $ being given by (\ref{sutodb}) and
\begin{equation}
{\cal H}\left({\xi_{j}^{*},\xi_{j-1}}\right)\equiv
{{\left\langle{\left.{\xi_{j}}\right\vert
H\left\vert{\xi_{j-1}}\right
\rangle}\right.}\over{\left\langle{{\xi_{j}}
\left\vert{\xi_{j-1}}\right\rangle}\right.}}\ .\label{sdiscaa}
\end{equation}
Here as in the introduction we have repeatedly inserted the
resolution
of unity (\ref{sutodb}) into the second relation  in
(\ref{tracecha}).
Within the trace (\ref{tracecha}), we always get a diagonalized
Hamiltonian by use of
$SU(2)$ rotation
\begin{equation}
UHU^\dagger =  h J_3\ , \quad   U \in SU(2),
\end{equation}
thus (\ref{tracecha}) is equivalent to the character formula;
\begin{equation}
Z(T) = {\rm Tr} \exp( -i hJ_3T ) =
 {\sin\left( (J+1/2) hT \right) \over \sin ( {hT/2} )}\
.\label{charact}
\end{equation}
Therefore, with the aid of (\ref{mele}), (\ref{sdisca}) is
found to be
\begin{equation}
 Z_N=  \prod\limits_{j=1}^N \int\limits_{\rm PBC} {d\mu \left(
{\xi_j^*,\xi_j}
\right)}
\exp \left[ iJ \left\{  2i\ln \left( {{1+\xi_j^*\xi_j} \over
{1+\xi_j^*
\xi_{j-1}} }\right) + \epsilon h
{{1 - \xi_j^*\xi_{j-1}} \over {1+\xi_j^*\xi_{j-1}} }
\right\} \right].\label{partd}
\end{equation}

In the following we shall consider the case that $J$ becomes large;
where the saddle point of the exponent in (\ref{partd}) is important,
which is given by a set of equations,
\begin{equation}
{\xi_j-\xi_{j-1}\over 1+\xi^*_j\xi_j}
={-i\epsilon h \xi_{j-1}\over 1+\xi^*_j\xi_{j-1}}\ ,\quad
{\xi^*_j-\xi^*_{j+1}\over 1+\xi^*_j\xi_j}
={-i\epsilon h \xi_{j+1}\over 1+\xi^*_{j+1}\xi_j}\quad
\left({1\le j\le N}\right)\ .\label{classceq}
\end{equation}
There are two solutions, satisfying PBC;
\begin{equation}
\xi_j^c = 0 \equiv \xi_c^{(+)}\ , \quad
\xi_j^c = \infty \equiv \xi_c^{(-)} \quad
\label{solutn}
\end{equation}
whose (finite!) contribution to the exponent (\ref{partd}) is read as
$e^{ihJT} (e^{-ihJT})$  for $\xi_c^{(+)} (\xi_c^{(-)})$.

By knowing that the measure $d\mu(\xi^*, \xi)$ in (\ref{sutodb}) is
invariant under $\xi \mapsto 1/ \xi$, and that $\prod_{j=1}^N( \xi_j
/
\xi_{j-1}) = 1$ under PBC, the change of variable, $\xi_j \mapsto 1/
\xi_j$, brings (\ref{partd}) to
\begin{equation}
 Z_N=  \prod\limits_{j=1}^N \int\limits_{\rm PBC}  {d\mu \left(
{\xi_j^*,\xi_j}
\right)}
\exp \left[ iJ \left\{  2i\ln \left( {{1+\xi_j^*\xi_j} \over
{1+\xi_j^*\xi_{j-1}} }
\right) - \epsilon h {{1 - \xi_j^*\xi_{j-1}} \over
{1+\xi_j^*\xi_{j-1}} }
\right\} \right].\label{anopar}
\end{equation}
Apparently the two expressions are related by $h \mapsto -h$.

\subsection{$SU(1,1)$;}
$su(1,1)$ algebra is given by
\begin{equation}
 [K_1, K_2] =-iK_3 , \quad [K_2, K_3] =iK_1,
\quad  [K_3, K_1] =iK_2,\label{noncsua}
\end{equation}
with
${\bf K}^2 = K_1^2 + K_2^2 - K_3^2 $, or
\begin{equation}
[K_+ ,K_-] =  2K_3 ,
\quad [K_3, K_\pm] =  \pm K_\pm; \quad
K_{\pm} \equiv \pm ( K_1 \pm iK_2 )
\  ;  ( K_+)^\dagger = -K_- ,\label{noncsub}
\end{equation}
with ${\bf K}^2 =  - K_+K_- - K_3^2 + K_3 = - K_-K_+ - K_3^2 - K_3 .$

We confine ourselves in a discrete series \cite{WB} to write
\begin{eqnarray}
 {\bf K}^2 |K, \ M\rangle & =  & K(1-K) |K, \ M\rangle,
 \quad  K_3|K, \ M\rangle = M|K, \ M\rangle   \nonumber\\
 K_{\pm} |K,  \ M \rangle &= & \pm \sqrt{(M\pm K)(M\mp K \pm 1)}
|K, M\pm 1\rangle , \label{noncsurep}\\
 M  &= &  K,\ K+1,\ K+2,\ \ldots \quad,
\quad   K = 1/2,\ 1,\ 3/2,\ 2,\ \ldots \quad.\nonumber
\end{eqnarray}
Adopting $|K, \ K \rangle$ as the fiducial vector,
$K_-|K, \ K \rangle = 0 ,$
as before, we obtain
\begin{eqnarray}
\left\vert{\xi}\right)  &\equiv&  e^{\xi K_{+}}\left\vert{K, K}
\right\rangle = \sum_{m=0}^{\infty}\xi^m {{2K+m-1}\choose
m}^{1/2}\left\vert{K, K+m}\right\rangle ,\nonumber \\
 \xi & \in & D_{(1,1)}= \left\{{\xi\in{\bf C},\ \vert\xi\vert<
1}\right\} . \label{suooa}
\end{eqnarray}
Their inner product is found to be
\begin{eqnarray}
\left({\xi\vert \xi^{\prime}}\right)
&=& \sum_{m=0}^{\infty} \left(\xi^{*}\xi^{\prime}\right)^{2m}
{{2K+m-1}\choose m}\nonumber\\
&=&\sum_{m=0}^{\infty} { \left(\xi^{*}\xi^{\prime}\right)^{2m} \over
m! } {1 \over \Gamma( 2K) } \int_0^\infty  dx x^{2K+m-1} e^{-x}
\label{normaa}\\
&=&
\left({1\over{1-\xi^{*}\xi^{\prime}}}\right)^{2K}\quad,
\nonumber
\end{eqnarray}
where we have introduced the integral representation of
$\Gamma$-function,
\begin{equation}
\int_0^\infty dx x^{n-1} e^{-x} = \Gamma(n) , \label{gammaf}
\end{equation}
in the second line, then performed
the summation with respect to
$m$ and  used (\ref{gammaf})  again in the final expression. Thus the
normalized state is given as
\begin{eqnarray}
\left\vert{\xi}\right\rangle & \equiv & {1\over{\left({\xi\vert
\xi}\right)^{1/2}}}\left\vert{\xi}\right) \nonumber \\
& =&\left({1-{\vert
\xi\vert}^{2}}\right)^{K}
\sum_{m=0}^{\infty} \xi^{m}{{2K+m-1}\choose m}^{1/2}\left\vert{K,\
K+m}\right\rangle\quad,\label{suooaa}
\end{eqnarray}
whose inner product and the resolution of unity in this case read as
\begin{mathletters}
\label{suood}
\begin{equation}
\left\langle{{\xi}\left\vert{\xi^{\prime}}\right\rangle}\right.=
{{\left({1-{\vert
\xi\vert}^{2}}\right)^{K}\left({1-{\vert
\xi^{\prime}\vert}^{2}}\right)^{K}}
\over{\left({1-\xi^{*}\xi^{\prime}}\right)^{2K}}}\ , \label{suood:a}
\end{equation}
\begin{equation}
{{2K-1}\over\pi}\int\limits_{D_{(1,1)}}
{{d\xi^{*}d\xi}\over{\left({1-|\xi|^2}\right)^{2}}}
\left\vert{\xi}\right\rangle\left\langle{\xi}\right\vert \equiv
\int\limits_{D_{(1,1)}} d\mu(\xi^{*},\xi)
\left\vert{\xi}\right\rangle\left\langle{\xi}\right\vert
 ={\bf 1}_{K}\quad,  \label{suood:b}
\end{equation}
\end{mathletters}
where
$ {\bf 1}_{K}\equiv\sum_{m=0}^{\infty}\left\vert{K,\ K+
m}\right\rangle\left\langle{K,\ K+m}\right\vert.
$ (It should be noted that  regularization for the
measure of K=1/2 is needed. See Appendix A.)

Matrix elements of generators are obtained as
\begin{eqnarray}
\left\langle {\left. \xi  \right|K_3\left| {\xi '} \right.}
\right\rangle &=&K{{1+\xi ^*\xi '} \over {1-\xi ^*\xi '}}
\left\langle {\xi }
\mathrel{\left | {\vphantom {\xi  {\xi '}}} \right.
\kern-\nulldelimiterspace}
{{\xi '}} \right\rangle\quad, \nonumber\\
  \left\langle {\left. \xi  \right|K_+\left| {\xi '} \right.}
\right\rangle
&=&K{{2\xi ^*} \over {1-\xi ^*\xi '}}\left\langle {\xi }
\mathrel{\left | {\vphantom {\xi  {\xi '}}} \right.
\kern-\nulldelimiterspace} {{\xi '}}
\right\rangle\quad. \label{matele}
\end{eqnarray}
Now we build up the path integral formula: again start with
the trace formula
\begin{eqnarray}
 Z(T)& =& {\rm Tr} e^{-iHT}  \nonumber\\
&=&\sum_{m=0}^\infty \langle K, \ K+m |  e^{-iHT} | K,
\ K+m \rangle\quad\hbox{(in ortho-normal basis)}\label{tracedisc}\\
&=&\int d\mu (\xi )\langle\xi | e^{-iHT}
|\xi\rangle
=\lim_{N \rightarrow\infty} Z_N
\quad\hbox{(in coherent basis)}\nonumber
\end{eqnarray}
with the Hamiltonian of the form
\begin{equation}
H=h_1K_1+h_2K_2-h_3K_3,\quad h_1^2+h_2^2-h_3^2 < 0
\end{equation}
which can be diagonalized as
\begin{equation}
V^{-1}HV = hK_3\ ,\quad V\in SU(1,1).
\end{equation}
Therefore $Z_{N}$ in (\ref{tracedisc}) is given by
\begin{equation}
Z_N = \prod\limits_{j=1}^N\int\limits_{\rm PBC} d\mu
\left(\xi_j^*,\xi_j \right) \exp
\left[ iK  \left\{  2i\ln \left(    {1-\xi_j^*\xi_{j-1}
\over {1-\xi_j^*\xi_j} } \right) -\epsilon h {1+
\xi_j^*\xi_{j-1} \over 1-\xi_j^*\xi_{j-1}}
\right\}  \right] . \label{dzpfin}
\end{equation}
It should be noted that $\xi $'s are in $D_{(1,1)}$ and
the spectrum in the trace formula (\ref{tracedisc}) is unbounded in
this
case while bounded in the $SU(2)$ case.

Now under $K$ being large, consider the saddle point conditions,
\begin{equation}
{\xi_j-\xi_{j-1}\over 1-\xi^*_j\xi_j}
={-i\epsilon h \xi_{j-1}\over 1-\xi^*_j\xi_{j-1}}\ ,\quad
{\xi^*_j-\xi^*_{j+1}\over 1-\xi^*_j\xi_j}
={-i\epsilon h \xi_{j+1}\over 1-\xi^*_{j+1}\xi_j}\quad
\left({1\le j\le N}\right)\quad .
\label{claseqdic}
\end{equation}
which has only one solution,
\begin{equation}
\xi_j^c = 0 \equiv \xi_c^{(0)}  ,
\label{soludisc}
\end{equation}
whose value of the exponent in (\ref{dzpfin}) is $e^{-ihKT}$, because
the other solution, $\xi = \infty$, for (\ref{claseqdic}) is outside
of
$D_{(1,1)}$.

\section{Exactness of the WKB approximation}

The WKB approximation is valid as the saddle point method when
$J(K)$ becomes large in $SU(2)$ $(SU(1,1))$ case.
To make the expansion transparent, let us put
\begin{equation}
\xi_j = \sqrt{\kappa} z_j\ , \label{wkba}
\end{equation}
in (\ref{partd}), (\ref{anopar}), and
(\ref{dzpfin}) respectively. Here
\begin{equation}
\kappa \equiv
\left\{{
\begin{array}{cccl}
{{\displaystyle 1} \over{\displaystyle 2J+1}}
& : & SU(2) &\\
{{\displaystyle 1} \over{\displaystyle 2K-1}}
& : & SU(1,1) &.\\
\end{array}
}\right. \label{param}
\end{equation}
It would be needless to say that the expression
(\ref{anopar}) is suitable when
expanding around $\xi = \infty$.  Plugging (\ref{wkba}) into
(\ref{partd}), (\ref{anopar}), and
(\ref{dzpfin}) and expanding the logarithm, we have
\begin{eqnarray}
Z_{N}^{(\alpha)}(\kappa^{(\alpha)})
& \equiv & e^{i h^{(\alpha)}J^{(\alpha)}T}
\prod_{j=1}^{N}\int\!{{{dz^{*}_{j}dz_{j}}\over\pi}
\exp\Biggl[{-(z^{*}_{j}z_{j}-e^{- i\epsilon h^{(\alpha)} }
z^{*}_{j}z_{j-1})}} \nonumber \\
\mbox{}& &+\sum_{n=1}^{\infty}{{{(-)^{n}}\over{n}}
(\kappa^{(\alpha)})^{n}\left[{
\left\{{\left({z^{*}_{j}z_{j}}\right)^{n} +\left({e^{- i\epsilon
h^{(\alpha)}}z^{*}_{j}z_{j-1}}\right)^{n}}\right\}}\right.}
\label{wkbab}\\
\mbox{}& &-\left.{{n\over{n+1}}\left\{{\left({z^{*}_{j}z_{j}}
\right)^{n+1}
-\left({e^{ -i\epsilon h^{(\alpha)}}
z^{*}_{j}z_{j-1}}\right)^{n+1}} \right\}}\right]\Biggr]\quad,
\nonumber
\end{eqnarray}
where we have discarded $O(\epsilon^2)$ terms to arrange the
expression and $\alpha$ takes the value $+$, $-$, and $0$ corresponding to
(\ref{partd}), (\ref{anopar}), and (\ref{dzpfin}) respectively.
(Recall the label of saddle point solutions;
(\ref{solutn}) and (\ref{soludisc}).) Thus
\begin{equation}
\left({h^{(\pm)} ,\  h^{(0)} }\right)\equiv
\left({ \pm h ,\  h }\right) ;\quad
\left({J^{(\pm)} , \ J^{(0)} }\right)\equiv
\left({ J ,\  K }\right) ;\quad
\left({\kappa^{(\pm)} ,\ \kappa^{(0)}}\right)\equiv
\left({\kappa , \  -\kappa }\right)\ .\label{a}
\end{equation}
In terms of $ Z_{N}^{(\alpha)}(\kappa^{(\alpha)})$, $Z_N$ is given, when
$J\ {\rm or}\ K\rightarrow \infty$,
\begin{eqnarray}
Z_N  & = & Z_{N}^{(+)}(\kappa^{(+)}) + Z_{N}^{(-)}(\kappa^{(-)})
\qquad  : \qquad  SU(2)\ , \nonumber \\
Z_N & = & Z_{N}^{(0)}(\kappa^{(0)})    \qquad \qquad \qquad
\qquad : \qquad SU(1,1)\ . \label{asmptz}
\end{eqnarray}
Therefore if the path integral in $SU(2)$ or $SU(1,1)$ would be
WKB-exact, the target is to prove the relation;
\begin{equation}
 Z_{N}(\kappa)  = Z_{N}(0) =e^{ihJT}
\prod_{j=1}^{N}\int\!{{{dz^{*}_{j}dz_{j}}\over\pi}}
\exp\left\{{-(z^{*}_{j}z_{j}-e^{-i\epsilon h}
z^{*}_{j}z_{j-1})}\right\}\ ,\label{target}
\end{equation}
for any $\alpha$. (Thus we have omitted the superscript $\alpha$.)
To this end, introduce parameters, $a_{j}$'s and $b_{j}$'s, then write
\begin{eqnarray}
Z_N(\kappa)
& = &\exp \left[ {\sum\limits_{n=1}^\infty  {{(-)^n}
\over n}\kappa^n\sum\limits_{j=1}^N
\left[ {\left\{ {\left( {-{\partial \over
{\partial a_j}}} \right)^n+\left( {{\partial  \over {\partial b_j}}}
\right)^n} \right\}}\right.}\right.\nonumber\\
\mbox{}& &-{\left.{\left.{
{n \over {n+1}}\left\{ {\left( {-{\partial  \over {\partial
a_j}}} \right)^{n+1}-\left( {{\partial  \over {\partial b_j}}}
\right)^{n+1}} \right\}} \right]} \right]} \nonumber\\
\mbox{}& &\times\prod\limits_{j=1}^N \int {  {dz_j^*dz_j \over \pi}
\exp \left\{ {- {\left( {a_jz_j^*z_j-  b_j e^{-i\epsilon
h}z_j^*z_{j-1}} \right)}} \right\}}\left. \right|_{\left\{a\right\}
=\left\{b\right\}=1} 
\label{prff} \\
&=&\prod_{i=1}^{N} F(\kappa; -\partial_{a_i},\partial_{b_i}) \left.
\int_0^\infty\! d\tau G\left(
({\bf a}-\gamma {\bf b})\tau  \right) \right|_{\left\{a\right\}
=\left\{b\right\}=1}\ ,\quad
\gamma \equiv e^{-ihT}\ ,\nonumber
\end{eqnarray}
where 
\begin{equation}
F\left(\kappa;x,y\right)\equiv
\exp\left[\sum^\infty_{n=1}{\left(-1\right)^n\over n}\kappa^n
\left\{\left( x^n+y^n\right)-{n\over n+1}\left( x^{n+1}-y^{n+1}\right)
\right\}\right]\ ,\label{fteig}
\end{equation}
and, for later convinience, we have introduced
\begin{equation}
G\left( ({\bf a}-\gamma{\bf b})\tau \right) \equiv \exp \left\{-\left(
{\bf a}-\gamma{\bf b}\right)\tau\right\} \ ;\label{gauss}
\end{equation}
since
\begin{eqnarray}
& &\prod\limits_{j=1}^N   \int  {dz_j^*dz_j \over \pi}
\exp \left\{ - \left( {a_jz_j^*z_j-e^{-i\epsilon h} b_jz_j^*z_{j-1}}
\right) \right\}
 =  { 1  \over {\prod\limits_{j=1}^N {a_j} -\gamma
\prod\limits_{j=1}^N {b_j}} } \equiv {1 \over {\bf a} - \gamma {\bf b}}
\label{bns}\\
& &=\int^\infty_0d\tau G\left(\left({\bf a}-\gamma{\bf b}\right)\tau\right)
\ .\nonumber
\end{eqnarray}
(Thus the condition, ${\rm Re}\left({\bf a}-\gamma{\bf b}\right)>0$,
should be understood.) 
Here $a$'s and $b$'s have been put unity after all, designating
$\left\{ a\right\}=\left\{ b\right\}=1\left(\equiv\left\{ a_1=b_1=\cdots
=a_N=b_N=1\right\}\right)$.

The important ingredient for the proof 
\clearpage
{\flushleft{\bf Formula;}}
\begin{equation}
F\left(\kappa;-\partial_{a_i},\partial_{b_i}\right)\left.\int^\infty_0
d\tau G\left(\left({\bf a}-\gamma{\bf
b}\right)\tau\right)\right\vert_{a_i=b_i=1}
=\int^\infty_0d\tau G\left(\left(\hat{\bf a}_i-\gamma\hat{\bf b}_i
\right)\tau\right)\ ,\label{fgg}
\end{equation}
where again ${\rm Re}\left(\hat{\bf a}_i-\gamma\hat{\bf b}_i\right)>0$
has been assumed.
Here and hereafter the carret designates the omission of the label such that
\begin{eqnarray}
\hat{\bf a}_i&\equiv& a_1a_2\cdots a_{i-1}a_{i+1}\cdots a_N=\prod_{j\ne i}a_j\
,
\nonumber\\
\hat{\bf a}_{i_1i_2}&\equiv& a_1\cdots a_{i_1-1}a_{i_1+1}\cdots
a_{i_2-1}a_{i_2+1}\cdots a_N=\prod_{j\ne i_1,i_2}a_j\ ,\label{tocar}\\
\hat{\left\{ a\right\}}_i&\equiv& \left\{ a\right\}\setminus a_i\ .\nonumber
\end{eqnarray}
The formula is then applied to the last line of (\ref{prff}) to give
\begin{eqnarray}
& &\prod^N_{i=1}F\left(\kappa;-\partial_{a_i},\partial_{b_i}\right)
\left.\int^\infty_0d\tau G\left(\left({\bf a}-\gamma{\bf b}\right)\tau\right)
\right\vert_{\left\{ a\right\}=\left\{ b\right\}=1}\\
&=&\prod_{i\ne j_1}F\left(\kappa;-\partial_{a_i},
\partial_{b_i}\right)\left.\int^\infty_0d\tau G\left(\left(\hat{\bf a}_{j_1}
-\gamma\hat{\bf b}_{j_1}\right)\tau\right)\right\vert_{\hat{\left\{ a\right\}
}_{j_1}=\hat{\left\{ b\right\}}_{j_1}=1}\ \ \left( 1\le{}^\forall j_1
\le N\right)\ ,\nonumber
\end{eqnarray}
which is further reduced to
\begin{equation}
=\prod_{i\ne j_1,j_2}F\left(\kappa;-\partial_{a_i},
\partial_{b_i}\right)\left.\int^\infty_0d\tau G\left(\left(\hat{\bf
a}_{j_1j_2}
-\gamma\hat{\bf b}_{j_1j_2}\right)\tau\right)\right\vert_{\hat{\left\{
a\right\}
}_{j_1j_2}=\hat{\left\{ b\right\}}_{j_1j_2}=1}\ .\nonumber\\
\end{equation}
Repeating these steps, we finally obtain the relation in which only
one $a$ and $b$ survive to find,
\begin{eqnarray}
&=&F\left(\kappa;-\partial_{\hat a_{j_1\cdots j_{N-1}}},
\partial_{\hat b_{j_1\cdots j_{N-1}}}\right)\left.\int^\infty_0
d\tau G\left(\left({\bf a}-\gamma{\bf b}\right)\tau\right)
\right\vert_{\hat a_{j_1\cdots j_{N-1}}=\hat b_{j_1\cdots j_{N-1}}=1}
\nonumber\\
&=&F\left(\kappa;-\partial_a,\partial_b\right)\left.\int^\infty_0
d\tau G\left(\left(a-\gamma b\right)\tau\right)\right\vert_{a=b=1}
\label{keisan}\\
&=&{1\over 1-\gamma}\ ,\nonumber
\end{eqnarray}
which proves the desired relation (\ref{target}) in view of (\ref{bns}).
Thus (\ref{target}) has been confirmed so that (\ref{asmptz}) is found to be
\begin{eqnarray}
Z_N & = &
{{e^{ihJT}}\over{1-e^{-ihT}}} + {{e^{-ihJT}}\over{1-e^{ihT}}}
= {\sin((J+1/2)hT) \over{\sin(hT/2)} }      \quad : \quad SU(2)\ ,
\nonumber\\
Z_N & =& {e^{-ihKT} \over{1-e^{-ihT}}}   \quad : \quad SU(1,1).
\label{fresu}
\end{eqnarray}
The conclusion is therefore that {\sl the WKB is exact under
coherent state path integrals for
$SU(2)$ and $SU(1,1)$(in the discrete series)}\/.

Now we proceed to establish the formula (\ref{fgg}): write 
\begin{eqnarray}
&F\left(\kappa;-\partial_{a_i},\partial_{b_i}\right)&
\left.\int^\infty_0d\tau G\left(\left({\bf a}-\gamma{\bf b}\right)\tau\right)
\right\vert_{a_i=b_i=1}\\
& &=\int^\infty_0d\tau F\left(\kappa;\hat{\bf a}_i\tau,\gamma\hat{\bf b}_i
\tau\right) G\left(\left(\hat{\bf a}_i-\gamma\hat{\bf b}_i\right)
\tau\right)\ ,\nonumber
\end{eqnarray}
so that the formula becomes 
\begin{equation}
\int^\infty_0d\tau F\left(\kappa;a\tau,\gamma b\tau\right) 
G\left(\left( a-\gamma b\right)\tau\right)
=\int^\infty_0d\tau G\left(\left(a-\gamma b\right)\tau\right)\ .
\label{fab}
\end{equation}
To make clear the $\kappa$-dependence of $F$, introduce
\begin{equation}
F\left(\kappa;x,y\right)=\sum^\infty_{n=0}{\kappa^n\over n!}
F_n\left( x,y\right)\ .
\end{equation}
First let us prove
\begin{equation}
\int^\infty_0d\tau F_n\left(\tau,c\tau\right) G\left(\left(1-c\right)\tau
\right)={\delta_{n,0}\over 1-c}\ ,Re(1-c)>0\ .
\label{fmone}
\end{equation}
$F_n$ is written explicitly as
\begin{eqnarray}
F_n(\tau, c\tau)&=&{n!\over 2\pi i}\oint\limits_{C_0}\!{dz\over z^{n+1}}
\exp\left[{\sum_{m=1}^{\infty}{{(-)^{m}}\over{m}}(\tau z)^{m}
\left\{{\left(1+c^{m}\right)
-{{m}\over{m+1}}\tau\left(1-c^{m+1}\right)}\right\}}\right]
\label{phint}\\
& = &{{n!}\over{2\pi i}}\oint\limits_{C_{0}}\!{{dz}\over{z^{n+1}}}
\exp\left\{ {-\ln(1+\tau z)(1+c\tau z) -{1\over z}\ln\left({{1+\tau
z}\over{1+c\tau z}}\right)+\left( 1-c\right)\tau}\right\}\
,\nonumber
\end{eqnarray}
where the contour $C_0$ is supposed to enclose the origin with
sufficiently small radius. Then (\ref{fmone}) becomes
\begin{eqnarray}
{\rm L.H.S. of \ (\ref{fmone})}  & \equiv &
\lim_{R\rightarrow\infty} \Omega_n(R)\ ,  \nonumber\\
\Omega_n(R) & = & {{n!}\over{2\pi i}}\oint\limits_{C_{R}}\!{{dz}
\over{z^{n+1}}}\int_{0}^{R}\!{{d\tau}\over{(1+\tau z)^{2}}}
\exp\left\{{-\left({{1\over z}-1}\right)
\ln\left({{1+\tau z}\over{1+c\tau z}}\right)}\right\}\quad,
\label{omegaa}
\end{eqnarray}
where $R$, permitting us to sum up the series to the logarithm in 
(\ref{phint}), is a cut-off and $C_R;\ |zR| <1$ which encloses the origin.
A change of variable
\begin{equation}
\tau\quad\mapsto\quad\rho
=\ln\left({{1+\tau z}\over{1+c \tau z}}\right), \quad
{{d\tau}\over{(1+\tau z)^{2}}}
={{d\rho}\over{(1-c )z}}e^{-\rho}\ ,
\label{taupho}
\end{equation}
brings us to
\begin{eqnarray}
\Omega_{n}(R) & = & {{n!}\over{2\pi
i}}\oint\limits_{C_{R}}\!{{dz}\over{z^{n+1}}}
\int_{0}^{\rho(R)}\!{{d\rho}\over{(1-c)z}}\exp\left({-{{\rho }
\over{z}}}\right)\nonumber\\
& = & {1\over{1-c}}\left({\delta_{n,\ 0}-\Psi_{n}(R)}\right)\quad,
\label{omegadelta}
\end{eqnarray}
where
\begin{eqnarray}
\Psi_{n}(R) & \equiv & {{n!}\over{2\pi
i}}\oint\limits_{C_{R}}\!{{dz}\over{z^{n+1}}}
\exp\left\{{-{1\over z}
\ln\left({{1+Rz}\over{1+c Rz}}\right)}\right\}\nonumber\\
& = & \left.\left({{d}\over{dz}}\right)^{n}
\exp\left\{{-{1\over z}\ln\left({{1+Rz}\over{1+c Rz}}\right)}\right\}
\right\vert_{z=0}\ ,
\label{psidiff}
\end{eqnarray}
$\Psi_{n}(R)$ satisfies a relation as;
\begin{eqnarray}
\Psi_{n}(R)
& = & \sum_{r=1}^{n}{{n-1}\choose{r-1}}\psi_{r}(R)\Psi_{n-r}(R)
,\qquad   \Psi_{0}(R) =e^{-(1-c)R}\ ,     \nonumber\\
\psi_{n}(R) & \equiv & {{(-)^n}\over{n+1}}n!R^{n+1}(1-c^{n+1})\ .
\label{psirec}
\end{eqnarray}
Apparently $\Psi_{n}(R)$ is found to be
\begin{equation}
\Psi_{n}(R)=(\hbox{polynomial of }R\hbox{ of order }2n)
\times e^{-(1-c)R} \label{psipol}
\end{equation}
and consequently
\begin{equation}
\lim_{R\rightarrow\infty}\Psi_{n}(R)=0\quad(n<\infty)\ .
\label{psilim}
\end{equation}
This completes the proof of (\ref{fmone}). Then with setting $\tau \mapsto
a\tau; c =\gamma b/a$, we see (\ref{fab}) holds. Thus we have proven
the formula (\ref{fgg}).

\section{Discussion}
The discussion in the previous section leads us to the fact that in
the
 path integral representation, (\ref{sutoc}) and (\ref{suooaa}), the
WKB
approximation is exact, whose result, as can been seen such as from
(\ref{keisan}), is independent of the number of time slices($N$).
Therefore putting $N \rightarrow \infty$ the result may be regarded
as
an infinite dimensional version of D-H theorem. To see the meaning of
this result clearly, let us re-examine the character formula
(\ref{partd}). First rewrite it with keeping the exponent always
correct
up to
$O(\epsilon)$ as
\begin{equation}
Z(T) =\lim_{N \rightarrow \infty} e^{ihJT}  \prod\limits_{j=1}^N
\int\limits_{\rm PBC} { {d\mu \left( {\xi _j^*,\xi _j} \right)}\exp
\left\{ {2J {\ln
\left( {{{1+e^{-i\epsilon h}\xi _j^*\xi _{j-1}} \over {1+\xi _j^*\xi
_j}}}
\right)}} \right\}}\quad,\label{zppfin}
\end{equation}
which becomes, by making use of local $U(1)$
invariance of the measure under
\begin{equation}
\xi_{j}\quad\mapsto\quad\xi_{j}e^{-ij\epsilon h}\quad,\label{utrns}
\end{equation}
to
\begin{eqnarray}
Z(T)
& = &\left.{\lim_{N \rightarrow \infty}
e^{ihJT}{\prod\limits_{j=1}^N}
\int d\mu \left( \xi_j^*,\xi_j \right)
\exp \left\{2J \ln \left({{1+ \xi_j^*\xi_{j-1}} \over
{1+\xi_j^*\xi_j}}\right)\right\}
}\right\vert_{\xi_{0}=\xi_{N}e^{-ihT}}\nonumber\\
& = & \left.{\lim_{N \rightarrow \infty}
e^{ihJT}{\prod\limits_{j=1}^N}
\int d\mu \left( \xi_j^*,\xi_j \right)
\langle \xi_j | \xi_{j-1} \rangle
}\right\vert_{\xi_{0}=\xi_{N}e^{-ihT}}
\label{zpfin}\\
& = &  e^{ihJT} \int  d\mu \left(\xi^*,\xi\right)
\langle \xi | \xi  e^{-ihT} \rangle\ ,\nonumber
\end{eqnarray}
where we have used (\ref{sutoda})
from the first to the second line and the resolution of unity
(\ref{sutodb}) from the second to the third line. After carrying out
the
trivial integration with respect to the phase of $\xi$, (\ref{zpfin})
reduces to
\begin{eqnarray}
Z(T)
& = & e^{ihJT}(2J+1)\int_{0}^{\infty}
{\!{{du}\over{(1+u)^2}}\left({{1+e^{-ihT}u}
\over{1+u}}\right)^{2J}}\nonumber\\
& = & e^{ihJT}\sum_{n=0}^{2J}e^{-inhT}
={{\sin((J+1/2)hT)}\over{\sin(hT/2)}}\quad.\label{zafin}
\end{eqnarray}

Let us take another point of view: make a change of variable
\begin{equation}
u\quad\mapsto\quad z=-\ln\left({{1+e^{-ihT}u}
\over{1+u}}\right)\quad,\label{vtrns}
\end{equation}
which brings the first line of (\ref{zafin}) to
\begin{equation}
Z(T) =
e^{ihJT}(2J+1)\int\limits_{C}\!{{dz}\over{1-e^{-ihT}}}e^{-(2J+1)z}\ .
\label{cntrint}
\end{equation}
Since the integrand in (\ref{cntrint}) has no singularity on
$z$-plane,
the original contour $C$ can be deformed arbitrarily. The new contour
$C_{+}+C_{R}+C_{-}$ illustrated in Fig.\ \ref{su2fig} provides an
interesting view point about the property of the character formula;
which reads
\begin{eqnarray}
Z(T)
& = &{{e^{ihJT}}\over{1-e^{-ihT}}}
\left\{{\int_{0}^{\infty}\!{dx}e^{-x}
-\int_{0}^{\infty}\!{dx}e^{-x-i(2J+1)hT}}\right\}\nonumber\\
& = &\left({{e^{ihJT}}\over{1-e^{-ihT}}} +
{{e^{-ihJT}}\over{1-e^{ihT}}} \right)
\int_{0}^{\infty}\!{dx}e^{-x} , \label{cntrinta}
\end{eqnarray}
giving the relation
\begin{eqnarray}
Z(T)
& = & {{\sin((J+1/2)hT)}\over{\sin(hT/2)}} =
{{e^{ihJT}}\over{1-e^{-ihT}}} + {{e^{-ihJT}}\over{1-e^{ihT}}}
\nonumber\\
& = &{\rm Tr} e^{-ih(a^\dagger a -J)T}
+ {\rm Tr} e^{ ih( a^\dagger a -J)T}\ .
\label{intrelat}
\end{eqnarray}
Hence it is
now clear that the character formula can be expressed as the sum of
partition functions of two harmonic oscillators with frequency $\pm
h$.

The situation is almost the same in $SU(1,1)$ case: corresponding to
(\ref{zppfin}), we have
\begin{equation}
Z(T) =\lim_{N \rightarrow \infty} e^{-ihKT}
\prod\limits_{j=1}^N
\int\limits_{\rm PBC} d\mu \left( \xi_j^*,\xi_j \right)
\exp\left\{ -2K \ln\left( {{1- e^{-i\epsilon h}\xi_j^*\xi_{j-1}}
\over{1-\xi_j^*\xi_j}}\right) \right\}\quad.\label{zdisc}
\end{equation}
By following the same procedure  from (\ref{zpfin}) to (\ref{zafin}),
$Z(T)$ can be calculated to be
\begin{eqnarray}
Z(T)
& = & e^{-ihKT}\int\!{d\mu \left( {\xi ^*,\xi } \right)}
\left\langle{\xi }\mathrel{\left | {\vphantom {\xi  {\xi e^{-ihT}}}}
\right.\kern-\nulldelimiterspace}
{{\xi e^{-ihT}}} \right\rangle \nonumber\\
& = &e^{-ihKT}(2K-1)\int_{0}^{1}
{\!{{du}\over{(1-u)^2}}\left({{1-u}
\over{1-e^{-ihT}u}}\right)^{2K}} \label{dzfin}\\
& = & e^{-ihKT}\sum_{n=0}^{\infty}e^{-inhT}
={{e^{-ihKT}}\over{1-e^{-ihT}}}\quad.\nonumber
\end{eqnarray}
Also similar from (\ref{vtrns}) to
(\ref{intrelat}), the final result (\ref{dzfin}) can be interpreted
as
follows: the change of variable,
\begin{equation}
u\quad\mapsto\quad z=-\ln\left({{1-u}
\over{1-e^{-ihT}u}}\right)\quad,\label{dtrns}
\end{equation}
gives us
\begin{equation}
Z(T) =e^{-ihKT}(2K-1)\int\limits_{C}\!{{dz}
\over{1-e^{-ihT}}}e^{-(2K-1)z}\ .\label{dcntrint}
\end{equation}
It is easily recognized from Fig.\ \ref{su1-1fig} that there exists
only one line integral which can be deformed to the line $0 \leq x <
\infty$, corresponding to the fact that there is only one harmonic
oscillator;
\begin{equation}
Z(T) = {{e^{-ihKT}}\over{1-e^{-ihT}}}
= {\rm Tr} e^{-ih(a^\dagger a + K)}\ . \label{interdic}
\end{equation}
The difference between contours in Fig.\ \ref{su2fig} and Fig.\
\ref{su1-1fig} is direct consequence of the difference between
compact
and non-compact phase spaces. This point of view will be also
developed
and become more transparent in the appendices.

Therefore we can see the reason why the WKB is exact;
however the situation is not always the case even in $SU(2)$ as was
mentioned in the introduction; if `the periodic coherent states
\cite{TK}',
\begin{eqnarray}
| \phi \rangle & \equiv & {1 \over \sqrt {2 \pi}} \sum_{M = -J}^{J}
e^{i(M+\Delta)\phi} | J, M\rangle ,\quad  -\pi \leq \phi< \pi,\
J=0,\ 1/2,\ 1,\ 3/2,\ \cdots\ ,\nonumber\\
\Delta & = & J-[J],\ [x]=\hbox{integer part of\ }x\ ,
\end{eqnarray}
is adopted then the final exponent is the Nielsen-Rohrlich
form\cite{NR};
\begin{eqnarray}
Z(T)
& = &{\rm Tr} e^{-iHT} = \lim_{N \rightarrow \infty}
\sum_{n=-\infty}^{\infty}
\int_{-\pi +2n\pi}^{\pi +2n\pi}d\phi_N
\prod_{k=1}^{N-1} \int_{-\infty}^{\infty} d\phi_k \nonumber\\
\mbox{}& & \times
\left. \int_{0}^{\pi} \lambda
\sin\theta_k d\theta_k \exp
\left\{ i(\lambda \cos \theta_k + \Delta )
(\phi_k-\phi_{k-1} )
- i \epsilon h \lambda \cos \theta_k   \right\}
\right |_{\phi_0 =\phi_N + 2\pi n}\ ,\label{exnr}
\end{eqnarray}
where $\lambda = J +1/2$. Naive continuum limit of (\ref{exnr})
matches
with that of (\ref{partd}) by putting
\begin{equation}
\xi = e^{i\phi} \tan{\theta \over 2}.
\label{xipt}
\end{equation}
The classical equations of motion, in terms of variables $\phi_k$
and $p_k\equiv \lambda\cos\theta_k$, are given by
\begin{equation}
\dot \phi \left( t \right) = h ,\quad \dot p \left( t \right) = 0 .
\end{equation}
Apparently there is no solution compatible to
the boundary condition,
$\phi\left( 0 \right) = \phi\left( T\right)  + 2n\pi$.
However in terms of $\phi_k$ and $\theta_k$,
\begin{equation}
\sin\theta\left( t\right)\left(\dot\phi\left( t\right) - h \right) =
0 ,\quad
\sin\theta\left( t\right)\dot\theta\left( t\right) = 0 .
\label{fsseq}
\end{equation}
There are solutions,
\begin{equation}
\theta = 0 \  {\rm or}\  \pi
\end{equation}
In view of (\ref{xipt}), $\theta = 0\left(\pi\right)$ corresponds to
$\xi_c = 0 \left( \infty \right)$ respectively, which tempts us build
up the relationship between our formula (\ref{partd}) and
(\ref{exnr}).
The task is now undertaken\cite{FKSF}.

The final comment is on the difference between (\ref{intrf}) and
(\ref{intrh}); namely if we had used the continuum path integral
formula
in (\ref{target}), we would get
$
 \sin{\left( JT\right)} /  \sin(T/2)
$
instead of the correct one;
${\sin\left(\left( J + 1/2\right)\right) T / \sin(T/2)}$.
The models which we have been considering are very much alike to
(three
dimensional) Chern-Simons theory \cite{WIT}. So the issue that
$J \rightarrow  J + 1/2$(Weyl shift) may correspond to the Coxeter
shift, $k \rightarrow k + 2$ (where $k$ denotes a level) in the
Chern-Simons case. Thus if it would be possible to perform the
integration in discretized version of the Chern-Simons theory, we
could
get the correct value $k+2$, which will be an interesting subject in
the
future.

Another task for us is such that owing to the technique in the
appendix
of obtaining the generalized coherent states from the canonical
coherent
one, we could generalize our discussion to the case of Grassmannian
manifold.

\appendix
\section{Coherent states from a geometrical view point}
We first summarize the
mathematical description (tensor product method\cite{RR}) of
constructing
coherent state for $SU(2)$ ($SU(1,1)$) as an example of compact
(non-compact) case in this appendix {\bf A}.
\subsection{$SU(2)$ case}
By parameterizing a point of
$
{\bf C}P^{1}
=\left\{{P\in M(2,{\bf C})|\ {\rm tr} P=1,
\quad P^{\dagger}=P,\quad P^{2}=P}\right\}
$
as
\begin{eqnarray}
P& = &{1 \over {1+\left| \xi \right|^2}}
\left( {\matrix{1&\xi^{*}\cr
\xi&{\left| \xi \right|^2}\cr }} \right)
={\hat u}(\xi){\hat u}^{\dagger}(\xi)\ ,\quad
{\hat u}(\xi)={{u(\xi)} \over{|u(\xi)|}}\nonumber\\
u(\xi)& = &\left( {\matrix{1\cr \xi\cr }} \right)
={\bf e}_{0}+\xi{\bf e}_{1}\in {\bf C}^{2}-\{0\}
\quad,
\label{appcpa}
\end{eqnarray}
$SU(2)$ system is described as a Hamiltonian system with symplectic
structure
\begin{eqnarray}
\omega & = & i{\rm tr}\left( {PdP\wedge dP} \right)\nonumber\\
& = & id{\hat u}^{\dagger}(\xi)\left({1-{\hat u}(\xi)
{\hat u}^{\dagger}(\xi)}\right)\wedge d{\hat u}(\xi)\label{appaa}\\
& = & i{{d\xi^{*}\wedge d\xi} \over
{\left( {1+\left| \xi \right|^2}
\right)^2}} \quad.\nonumber
\end{eqnarray}
The vectors ${\bf e}_{0},\ {\bf e}_{1}$ in (\ref{appcpa})
are basis vectors of ${\bf C}^{2}$ being $2$-dimensional
representation
space of $SU(2)$.
Our convention on this basis is
\begin{eqnarray}
d\rho_{1/2}(J_{3}){\bf e}_m
& = & (-1/2+m){\bf e}_m\ ,\nonumber\\
d\rho_{1/2}(J_{+}){\bf e}_m
& = & \sqrt{(m+1)(1-m)}{\bf e}_{m+1}\ ,\quad(\ m=0,1\ )\\
d\rho_{1/2}(J_{-}){\bf e}_m
& = & \sqrt{m(2-m)}{\bf e}_{m-1}\ .\nonumber
\end{eqnarray}
The dynamical variables are elements of
$su(2)$ being realized as functions on ${\bf C}P^{1}$ by a map
\begin{eqnarray}
su(2)\ni X\mapsto F_{X}(P)
& = & {\rm tr}\left( {Pd\rho_{1/2}(X)} \right) ={\rm tr}\left( {{\hat
u}(\xi) {\hat u}^{\dagger}(\xi) d\rho_{1/2}(X)} \right)\nonumber\\
& = & {\hat u}^{\dagger}(\xi)d\rho_{1/2}(X){\hat u}(\xi)\in {\bf R}\
{}.
\label{appab}
\end{eqnarray}
The Poisson bracket between two variables $X,\ Y\in su(2)$ is
defined in terms of their corresponding vector fields $V_{X},\ V_{Y}$
and $\omega^{-1}$ by
\begin{equation}
\left\{{F_{X},F_{Y}}\right\}_{\rm P.B.}\equiv\omega^{-1}
\left({V_{X},V_{Y}}\right)\ .\label{appcpb}
\end{equation}
Taking $J_{3}$ as a Hamiltonian, we find that the classical
action is given by
\begin{equation}
S={1 \over 2}\int_0^T {dt\left( {i{{\xi^{*}\dot \xi
-{\dot \xi}^{*}\xi}
\over {1+\left| \xi \right|^2}}
+{{1-\left| \xi \right|^2} \over
{1+\left| \xi \right|^2}}}
\right)}\quad.\label{appcpc}
\end{equation}
There are various symplectic structures corresponding to
higher spin representation. They are given by an embedding
\begin{equation}
{\bf C}P^{1}\rightarrow {\bf C}P^{2J}=\left\{{P\in M(2J+1,{\bf C}) |\
{\rm tr} P=1,\ P^{\dagger}=P,\ P^{2}=P}\right\} \ .\label{appcpe}
\end{equation}
To see this explicitly, we first note the well known fact that the
symmetric
sector of the tensor product $\otimes^{2J}{\bf C}^{2}$ is invariant
under the action of $(\otimes^{2J}\rho_{1/2})(g),\ g\in GL(2,{\bf
C})$. In
order to pick up this subspace, $V_{J}$, define a generating function
of basis vectors of $V_{J}$:
\begin{eqnarray}
V_{J} \ni  {\cal G}_{J}(t)
& = & \rho_{J}(e^{tJ_{+}})\left({{\bf e}_{0}
\otimes\cdots\otimes{\bf e}_{0}}\right)\nonumber\\
& = & \sum\limits_{m=0}^{2J}{
t^{m}\sum_{|{\bf n}|=m}{\bf e}_{n_{1}}\otimes{\bf e}_{n_{2}}
\otimes\cdots\otimes{\bf e}_{n_{2J}}},\label{cpogen}\\
|{\bf n}| & = & \sum_{i=1}^{2J}n_{i},\quad n_{i}=0,1\ ,\nonumber
\end{eqnarray}
where
\begin{equation}
\rho_{J}(g)v\equiv(\otimes^{2J}\rho_{1/2})(g)v,\quad
g\in GL(2,{\bf C}),\ v\in V_{J}\ .\label{restsutwo}
\end{equation}
Then we can find a set of basis vectors for $V_{J}$:
\begin{eqnarray}
{\bf e}_m^{\left( J \right)}
&=&\left( {\matrix{{2J}\cr m\cr }}
\right)^{-{1 \mathord{\left/ {\vphantom {1 2}} \right.
\kern-\nulldelimiterspace} 2}}
\left.{{1\over{m!}}\left({d\over{dt}}\right)^{m}}
\right\vert_{t=0}
{\cal G}_{J}(t) \ (0\le m\le 2J)\nonumber\\
&=&\left( {\matrix{{2J}\cr m\cr }}
\right)^{-{1 \mathord{\left/ {\vphantom {1 2}} \right.
\kern-\nulldelimiterspace} 2}}
\sum_{|{\bf n}|=m}{\bf e}_{n_{1}}\otimes{\bf e}_{n_{2}}
\otimes\cdots\otimes{\bf e}_{n_{2J}},\label{appcpg}\\
& &\left({{\bf e}_m^{\left( J \right)}
, {\bf e}_n^{\left( J \right)}}\right)
= \delta_{m,n}\ .\nonumber
\end{eqnarray}
The spin-$J$ representation of $su(2)$ on $V_{J}$ is expressed, by
putting
$g=e^{tX},\ X\in su(2)$ in (\ref{restsutwo}), as
\begin{equation}
d\rho _J\left( X \right)=\sum\limits_{k=1}^{2J}
{\overbrace {\rho _{1/ 2}\left(
1 \right)\otimes \cdots \otimes \rho _{1/ 2}
\left( 1 \right)}^{k-1}\otimes
d\rho _{1/ 2}\left( X \right)\otimes
\overbrace {\rho _{1/ 2}\left(
1 \right)\otimes \cdots \otimes \rho _{1/ 2}
\left( 1 \right)}^{2J-k}}
\ .\label{drhoX}
\end{equation}
In particular, corresponding to (\ref{sutwoalg}), we have
\begin{eqnarray}
d\rho_{J}(J_{3}){\bf e}_m^{\left( J \right)}
& = & (-J+m){\bf e}_m^{\left( J \right)}\ ,\nonumber\\
d\rho_{J}(J_{+}){\bf e}_m^{\left( J \right)}
& = & \sqrt{(m+1)(2J-m)}{\bf e}_{m+1}^{\left( J \right)}\ ,
\quad(0\le m\le 2J)\\
d\rho_{J}(J_{-}){\bf e}_m^{\left( J \right)}
& = & \sqrt{m(2J-m+1)}{\bf e}_{m-1}^{\left( J \right)}\ .\nonumber
\end{eqnarray}
Now defining
\begin{eqnarray}
u(\xi)& \mapsto & u_{J}(\xi)
={\cal G}_{J}\left({\xi}\right)\nonumber\\
& = & \sum\limits_{m=0}^{2J}
{\left( {\matrix{{2J}\cr m\cr }} \right)
^{{1 \mathord{\left/ {\vphantom {1 2}} \right.
\kern-\nulldelimiterspace} 2}}
\xi^m{\bf e}_m^{\left( J \right)}}
\in{\bf C}^{2J+1}-\{0\}\ , \quad
{\hat u}_{J}(\xi)={{u_{J}(\xi)}\over{|u_{J}(\xi)|}}\ ,\label{appcph}
\end{eqnarray}
we find an explicit parameterization of the embedding (\ref{appcpe}):
\begin{equation}
P_{J}={\hat u}_{J}(\xi){\hat u}_{J}^{\dagger}(\xi)
\in {\bf C}P^{2J}\ .\label{appcpd}
\end{equation}

Again the dynamical variables are given by
\begin{equation}
F_{X}^{(J)}(P)={\rm tr}\left( {P_{J}d\rho_{J}(X)} \right)
={\hat u}_{J}^{\dagger}(\xi)d\rho_{J}(X) {\hat u}_{J}(\xi)\ ,
\label{appcpi}
\end{equation}
with Poisson brackets being defined
with respect to the symplectic form
\begin{eqnarray}
\omega_{J} & = & i{\rm tr}\left( {P_{J}dP_{J}\wedge dP_{J}}\right)
\nonumber\\
& = & id{\hat u}_{J}^{\dagger}(\xi)\left({1-{\hat u}_{J}(\xi)
{\hat u}_{J}^{\dagger}(\xi)}\right)\wedge d{\hat u}_{J}(\xi)
\label{appcpj}\\
& = & 2iJ{{d\xi^{*}\wedge d\xi} \over
{\left( {1+\left| \xi\right|^2}\right)^2}} \quad.\nonumber
\end{eqnarray}
The classical action is read as
\begin{equation}
S_{J}=J\int_0^T {dt\left( {i{{\xi^{*}\dot \xi
-{\dot \xi}^{*}\xi}
\over {1+\left| \xi \right|^2}}+{{1-\left| \xi \right|^2}
\over {1+\left| \xi\right|^2}}} \right)}\quad.
\label{appcpk}
\end{equation}
Thus one might expect the path integral expression
\begin{equation}
\int {\prod\limits_{0\le t\le T} {\omega _J\left( t \right)}
\exp \left(
{iJ\int_0^T {dt\left\{ {i{{\xi^{*}\dot \xi-{\dot \xi}^{*}\xi}
\over{1+\left| \xi\right|^2}}
+{{1-\left| \xi \right|^2}
\over {1+\left| \xi \right|^2}}}
\right\}}} \right)}\label{appcpl}
\end{equation}
could correspond to the trace formula
${\rm Tr}\left({e^{-iJ_{3}T}}\right)$
with a suitable boundary condition. Such an expectation is, however,
too
naive and (\ref{appcpl}) lacks much information as was mentioned
in the introduction. As a final comment, it should be noted that the
vector space
$V_{J}$ spanned by
$\left\{{{\bf e}_{m}^{(J)};0\le m\le 2J}\right\}$ is just the $2J+1$
dimensional representation space of
$SU(2)$ and a coherent state in this representation, $u_{J}(\xi)$,
given by a symmetric tensor product, is nothing but a generating
function of the basis vectors of $V_{J}$.

\subsection{$SU(1,1)$ case}
The phase space $D_{(1,1)}$ for $SU(1,1)$ system is given by
\begin{eqnarray}
D_{(1,1)}& \equiv & \left\{{\xi\in {\bf C}|\
\left|{\xi}\right|^{2}<1}\right\}\nonumber\\
& = & \left\{{P\in M(2,{\bf C})|\
{\rm tr} P=1,\ \eta P^{\dagger}\eta=P,\ P^{2}=P}\right\}\quad,
\label{appdskaa}
\end{eqnarray}
where $\eta={\rm diag(1,-1)}$. An explicit form of $P$ is
\begin{eqnarray}
P& = &v(\xi)\left({\eta v(\xi)}\right)^{\dagger}
=v(\xi)v(\xi)^{\dagger}\eta\nonumber\\
& = & {1 \over {1-\left| \xi \right|^2}}
\left( {\matrix{1&-\xi^{*}\cr
\xi&-{\left| \xi \right|^2}\cr }} \right)\quad,
\label{appdskab}
\end{eqnarray}
with
\begin{equation}
v(\xi)={1\over {\sqrt{1-|\xi|^{2}}}}
\left( {\matrix{1\cr \xi\cr }} \right),\quad
v(\xi)^{\dagger}\eta v(\xi)=1\quad.\label{appdskac}
\end{equation}
The symplectic form on this phase space is found, as (\ref{appaa}),
to
be
\begin{equation}
\omega =i{\rm tr}\left( {PdP\wedge dP} \right)
=-i{{d\xi^{*}\wedge d\xi} \over
{\left( {1-\left| \xi \right|^2} \right)^2}} \quad.
\label{appdskad}
\end{equation}
Again similar to (\ref{appab}) the map
\begin{equation}
su(1,1)\ni X\mapsto F_{X}(P)={\rm tr}(PX)=v(\xi)^{\dagger}
\eta Xv(\xi)\quad,\label{appdskaf}
\end{equation}
gives us dynamical variables and classical mechanics on this phase
space
are described with symplectic form (\ref{appdskad}). Various
symplectic
structures are obtained by a similar recipe as above. However they do
not
provide unitary representation of $SU(1,1)$ due to the indefinite
metric
$\eta$.

To obtain the discrete series of
unitary representation of $SU(1,1)$ as well as coherent states,
an embedding $D_{(1,1)}\rightarrow{\bf C}P^{\infty}$ is needed: first
construct the fundamental representation ({\it limit of discrete
series\/}) as
\begin{eqnarray}
d{\tilde {\rho}}_{1/2}\left(K_{3}\right)& = &
\sum_{n=0}^{\infty}(n+1/2)
{\bf e}_{n}{\bf e}_{n}^{T}\nonumber\\
d{\tilde {\rho}}_{1/2}\left(K_{+}\right)& = &
\sum_{n=0}^{\infty}(n+1)
{\bf e}_{n+1}{\bf e}_{n}^{T}\label{appdska}\\
d{\tilde {\rho}}_{1/2}\left(K_{-}\right)& = &
-\sum_{n=0}^{\infty}(n+1)
{\bf e}_{n}{\bf e}_{n+1}^{T}\nonumber\\
\left({{\bf e}_{m},{\bf e}_{n}}\right)& = &{\bf e}_{m}^{T}{\bf e}_{n}
=\delta_{m,n}\nonumber
\end{eqnarray}
and the coherent state; by a map
\begin{equation}
D_{(1,1)}\ni \xi\mapsto u(\xi)=\sum_{n=0}^{\infty}\xi^{n}{\bf
e}_{n}\in l^{2}({\bf C})-\{0\}\quad,
{\hat u}(\xi)={{u(\xi)} \over{|u(\xi)|}}\ .\label{appdskb}
\end{equation}
Here ${\tilde {\rho}}_{1/2}$ is
\begin{equation}
{\tilde {\rho}}_{1/2}: SL(2,{\bf C})\ni g\ \mapsto\
{\tilde {\rho}}_{1/2}(g)\in GL(l^{2}({\bf C}))
\end{equation}
and ${\bf e}_{n}$'s are basis vectors of $l^{2}({\bf C})$. Formal
definition of
$e^{tK_{+}}$ is given by
\begin{equation}
{\tilde
{\rho}}_{1/2}(e^{tK_{+}})=\sum_{m=0}^{\infty}\sum_{n=0}^{\infty}
{t^{m}
\left( {\matrix{{m+n}\cr m\cr }} \right)
{\bf e}_{n+m}{\bf e}_{n}^{T}}\ ,\label{rhokplus}
\end{equation}
which generates a coherent state, when acting on ${\bf e}_{0}$:
\begin{equation}
{\tilde {\rho}}_{1/2}(e^{tK_{+}}){\bf e}_{0}=u(t),\ |t|<1 \
.\label{gencoh}
\end{equation}
The resolution of unity is written as
\begin{equation}
\int\limits_{D_{(1,1)}} d\mu(\xi){\hat u}(\xi){\hat u}(\xi)^{\dagger}
={\tilde {\rho}}_{1/2}\left(1\right)
=\sum_{n=0}^{\infty}{\bf e}_{n}{\bf e}_{n}^{T}\ ,
\label{appdskc}
\end{equation}
where we have introduced a regularized measure $d\mu(\xi)$ ,
\begin{equation}
d\mu(\xi)=\mathop{\lim }\limits_{\epsilon\rightarrow +0}
{\epsilon\over\pi} {{d\xi^{*} d\xi}
\over {\left( {1-\left| \xi \right|^2}
\right)^{2-\epsilon}}}\ .\label{appdskd}
\end{equation}
The remainder of the discrete series are
obtained by
\begin{equation}
D_{(1,1)}\ni \xi\rightarrow
P_{K}={\hat u}_{K}(\xi){\hat u}_{K}^{\dagger}(\xi)
\in{\bf C}P^{\infty}\quad,\label{appdske}
\end{equation}
where ${\hat u}_{K}(\xi)$ is given by the same procedure in the
previous section: first, define a generating function of basis
vectors
for symmetric sector (${\tilde V}_{K}$) of $\otimes^{2K}l^{2}({\bf
C})$
\begin{eqnarray}
{\tilde V}_{K} \ni  {\tilde {\cal G}}_{K}(t)
& = & {\tilde {\rho}}_{K}(e^{tK_{+}})
({\bf e}_{0}\otimes\cdots\otimes{\bf e}_{0})
\ (|t|<1)\nonumber\\
& = & \sum\limits_{m=0}^{\infty}{
t^{m}\sum_{|{\bf n}|=m}{\bf e}_{n_{1}}\otimes{\bf e}_{n_{2}}
\otimes\cdots\otimes{\bf e}_{n_{2K}}},\label{doogen}\\
|{\bf n}| & = & \sum_{i=1}^{2K}n_{i},\quad
n_{i}=0,\ 1,\ 2,\ \cdots ,\nonumber
\end{eqnarray}
where
\begin{equation}
{\tilde {\rho}}_{K}(g){\tilde v}
=(\otimes^{2K}{\tilde {\rho}}_{1/2})(g){\tilde v}\quad
g\in SL(2,{\bf C}),\ {\tilde v}\in {\tilde V}_{K}\ .\label{restsuo}
\end{equation}
With a similar manner as above a set of basis vectors for ${\tilde
V}_{K}$;
\begin{eqnarray}
{\bf e}_m^{\left( K \right)}
&=&\left( {\matrix{{2K+m-1}\cr m\cr }}
\right)^{-{1 \mathord{\left/ {\vphantom {1 2}} \right.
\kern-\nulldelimiterspace} 2}}
\left.{{1\over{m!}}\left({d\over{dt}}\right)^{m}}
\right\vert_{t=0}
{\tilde {\cal G}}_{K}(t)\label{doobase}\ (m=0,\ 1,\ 2,\ \cdots)
\nonumber\\
&=&\left( {\matrix{{2K+m-1}\cr m\cr }}
\right)^{-{1 \mathord{\left/ {\vphantom {1 2}} \right.
\kern-\nulldelimiterspace} 2}}
\sum_{|{\bf n}|=m}{\bf e}_{n_{1}}\otimes{\bf e}_{n_{2}}
\otimes\cdots\otimes{\bf e}_{n_{2K}},\label{doobaseb}\\
& &\left({{\bf e}_m^{\left( K \right)}
, {\bf e}_n^{\left( K \right)}}\right)
= \delta_{m,n}\nonumber
\end{eqnarray}
is found. Thus ``spin"-$K$ representation of $X\in su(1,1)$ is
expressed as
\begin{equation}
d{\tilde {\rho}} _K\left( X \right)=
\sum\limits_{k=1}^{2K} {\overbrace {{\tilde {\rho}}
_{1/ 2}\left( 1 \right)\otimes \cdots \otimes
{\tilde {\rho}} _{1/ 2}\left( 1 \right)}^{k-1}
\otimes d{\tilde {\rho}} _{1/ 2}\left( X \right)
\otimes \overbrace {{\tilde {\rho}}
_{1/ 2}\left( 1 \right)\otimes \cdots \otimes
{\tilde {\rho}} _{1/ 2}\left( 1 \right)}^{2K-k}}\ ,
\label{appdskf}
\end{equation}
yielding (corresponding to (\ref{noncsurep}))
\begin{eqnarray}
d{\tilde {\rho}}_{K}(K_{3}){\bf e}_m^{\left( K \right)}
& = & (K+m){\bf e}_m^{\left( K \right)}\ ,\nonumber\\
d{\tilde {\rho}}_{K}(K_{+}){\bf e}_m^{\left( K \right)}
& = & \sqrt{(m+1)(2K+m)}{\bf e}_{m+1}^{\left( K \right)}\ ,\\
d{\tilde {\rho}}_{K}(K_{-}){\bf e}_m^{\left( K \right)}
& = & -\sqrt{m(2K+m-1)}{\bf e}_{m-1}^{\left( K \right)}\ .\nonumber
\end{eqnarray}
We now define $\hat u_K\left( \xi \right)$ as
\begin{eqnarray}
u_K\left( \xi \right)
& = & {\tilde {\cal G}}_{K}(\xi)\nonumber\\
& = & \sum\limits_{m=0}^\infty
{\left( {\matrix{{2K+m-1}\cr m\cr }}
\right)^{{{1} \mathord{\left/ {\vphantom {{-1} 2}} \right.
\kern-\nulldelimiterspace} 2}}\xi^m{\bf e}_m^{\left( K \right)}}
\in l^{2}({\bf C})-\{0\}\ ,\quad
\hat u_K\left( \xi \right)={{u_K(\xi)}\over{|u_K(\xi)|}}\ ,
\label{appdski}
\end{eqnarray}
giving an explicit parameterization of $P_{K}$ in (\ref{appdske}).
Integrating $P_{K}$ by use of the measure
\begin{equation}
d\mu_{K}\left({\xi}\right)= {{2K-1} \over \pi}
{{d\xi^{*}d\xi}
\over {\left( {1-\left| \xi \right|^2}\right)^2}}\ ,
\label{appdskj}
\end{equation}
we find the resolution of unity
\begin{equation}
\int\limits_{D(1,1)} {d\mu _K\left( \xi \right)P_K
={\tilde {\rho}} _K\left( 1 \right)=\sum\limits_{m=0}^\infty
{{\bf e}_m^{\left( K \right)}
{{\bf e}_m^{\left( K \right)}}^T}}\ .\label{appdskk}
\end{equation}
In view of (\ref{doogen}) and (\ref{appdski}), we again recognize
that a coherent
state in ``spin"-$K$ representation is nothing but a generating
function of basis
vectors of the representation space ${\tilde V}_{K}$.

\section{Coherent states in terms of\protect\\ the Schwinger bosons}
\subsection{$SU(2)$ case}
Following Schwinger\cite{SB}, first consider a system of two bosonic
oscillators:
\begin{equation}
[a_{i},a^{\dagger}_{j}]=\delta_{i,j},\quad[a_{i},a_{j}]=0,
\quad[a_{i}^{\dagger},a^{\dagger}_{j}]=0\quad(i=1,\ 2)\ ,
\label{osca}
\end{equation}
and define operators
\begin{equation}
{\hat J}_{3}\equiv{\bf a}^{\dagger}{{\sigma_{3}}\over 2}
{\bf a},\quad {\hat J}_{\pm}
\equiv{\bf a}^{\dagger}{{\sigma_{1}\pm
i\sigma_{2}}\over 2}{\bf a}\label{oscb}
\end{equation}
The Fock space is given as usual:
\begin{eqnarray}
\left\vert{(n_1,n_2)}\right\rangle
&\equiv&{1\over{\sqrt{n_{1}!n_{2}!}}}
\left({a^{\dagger}_{1}}\right)^{n_{1}}
\left({a^{\dagger}_{2}}\right)^{n_{2}}
\left\vert{(0,0)}\right\rangle\quad
(n_{i}=0,\ 1,\ 2\ \cdots\ )\nonumber\\
\mbox{}& &a_{i}\left\vert{(0,0)}\right\rangle=0\
(i=1,\ 2)\quad.
\label{oscc}
\end{eqnarray}
Introducing a constraint,
\begin{equation}
{\bf a}^{\dagger}{\bf a}-2J=0\
\left(J=0,\ 1/2,\ 1,\ 3/2,\ \cdots\right)\ ,
\label{apposca}
\end{equation}
we can build up the spin-$J$
representation of $su(2)$ out of the Fock space:
\begin{eqnarray}
\left\vert{(J+M,J-M)}\right\rangle
& = &{1\over{\sqrt{(J+M)!(J-M)!}}}
\left({a^{\dagger}_{1}}\right)^{J+M}
\left({a^{\dagger}_{2}}\right)^{J-M}
\left\vert{(0,0)}\right\rangle\nonumber\\
& \equiv& \left\vert{J,M}\right\rangle\ .
\label{oscd}
\end{eqnarray}
Thus the $SU(2)$ coherent state is identified with
\begin{equation}
\left\vert{\xi}\right\rangle\equiv
{1\over{\left({1+|\xi|^{2}}\right)^{J}}}
\exp\left({\xi
a^{\dagger}_{1}a_{2}}\right)
\left\vert{(0,2J)}\right\rangle,
\quad\xi\in{\bf C}\ .\label{osce}
\end{equation}
The constraint (\ref{apposca}) is realized by a projection operator
in
the Fock space
\footnote{Rigorously speaking there need a regularization such that
\[
{\bf I}_{J}=\lim\limits_{\varepsilon\rightarrow +0}
\int_{0}^{2\pi}{{d\lambda}\over{2\pi}}
\exp\left\{{i\lambda\left({{\bf a}^{\dagger}
{\bf a}-2J}\right)-\varepsilon{\bf a}^{\dagger}{\bf a}
}\right\}
\]
in order to exchange the order of $\lambda$-integration and other
operations in a safe manner. (See (\ref{trosca})).
};
\begin{equation}
{\bf I}_{J}\equiv\int_{0}^{2\pi}{{d\lambda}\over{2\pi}}
\exp\left\{{i\lambda\left({{\bf a}^{\dagger}
{\bf a}-2J}\right)}\right\}\quad,\label{oscf}
\end{equation}
which is rewritten by use of canonical coherent state as
\begin{eqnarray}
{\bf I}_{J} & = & \int{{\left({dz^{*} dz}\right)^{2}}\over{\pi^{2}}}
{{\left({d{z^{\prime}}^{*} dz^{\prime}}\right)^{2}}\over{\pi^{2}}}
\left\vert{{\bf z}^{\prime}}\right\rangle
\left\langle{{\bf z}^{\prime}}\right\vert
\int_{0}^{2\pi}{{d\lambda}\over{2\pi}}
\exp\left\{{i\lambda\left({{\bf a}^{\dagger}{\bf
a}-2J}\right)}\right\}
\left\vert{{\bf z}}\right\rangle\left\langle{{\bf z}}\right\vert
\nonumber\\
& = &{1\over{(2J)!}}\int{{\left({dz^{*}
dz}\right)^{2}}\over{\pi^{2}}}
\left\vert{{\bf z}}\right\rangle\left\langle{(0,0)}\right\vert
\left({{\bf z}^{\dagger}{\bf a}}\right)^{2J}
\exp\left({-{1\over 2}{\bf z}^{\dagger}{\bf z}}\right)\quad.
\label{oscg}
\end{eqnarray}
Making a change of variables
\begin{equation}
\pmatrix{\displaystyle{z_{1}}\cr
\displaystyle{z_{2}}}=\zeta\pmatrix{\displaystyle{\xi}\cr
\displaystyle{1}},\quad \left({d z^{*}d z}\right)^{2}=|\zeta|^{2}
d\zeta^{*}d\zeta d\xi^{*}d\xi\quad,\label{osch}
\end{equation}
and integrating with respect to $\zeta$, we find
\begin{eqnarray}
{\bf I}_{J}& = &{1\over{(2J)!}}\int{{|\zeta|^{2} d\zeta^{*}d\zeta
d\xi^{*}d\xi}\over{\pi^{2}}} e^{\zeta\left(\xi
a_{1}^{\dagger}+a_{2}^{\dagger}\right)}
\left\vert{(0,0)}\right\rangle\left\langle{(0,0)}\right\vert
\left\{{\zeta^{*}\left(\xi^{*} a_{1}+a_{2}\right)}\right\}^{2J}
e^{-|\zeta|^{2}\left({1+|\xi|^{2}}\right)}\nonumber\\
& = &{{2J+1}\over{\pi}}\int{{d\xi^{*}d\xi}
\over{\left({1+|\xi|^{2}}\right)^{2}}}
{{\left(\xi a_{1}^{\dagger}+a_{2}^{\dagger}\right)^{2J}}
\over{\sqrt{(2J)!}\left({1+|\xi|^{2}}\right)^{J}}}
\left\vert{(0,0)}\right\rangle\left\langle{(0,0)}
\right\vert {{\Bigl(\xi^{*}
a_{1}+a_{2}\Bigr)^{2J}}
\over{\sqrt{(2J)!}\left({1+|\xi|^{2}}\right)^{J}}}\label{osci}\\
& = &\sum\limits_{m,n=0}^\infty  {{{2J+1} \over \pi }
\int {{{d\xi ^*d\xi } \over
{\left( {1+\left| \xi  \right|^2} \right)^2}}
{1 \over {\left( {1+\left| \xi
\right|^2} \right)^{2J}}}
{{\left( {\xi a_1^\dagger a_2} \right)^m} \over {m!}}\left|
{\left( {0,2J} \right)} \right\rangle \left\langle
{\left( {0,2J} \right)}
\right|{{\left( {\xi ^*a_1a_2^\dagger } \right)^n}
\over {n!}}}}\nonumber\\
& = &{{2J+1}\over{\pi}}\int{{d\xi^{*}d\xi}
\over{\left({1+|\xi|^{2}}\right)^{2}}}
\left\vert{\xi}\right\rangle\left\langle{\xi}\right\vert
\quad.\nonumber
\end{eqnarray}
Thus we see the equivalence between the projection operator
${\bf I}_{J}$ on the Fock space and the resolution of unity in the
spin-$J$
representation of $SU(2)$.

There is another solution for constraint (\ref{apposca})
\begin{equation}
\pmatrix{\displaystyle{z_{1}}\cr
\displaystyle{z_{2}}}=i\zeta\pmatrix{\displaystyle{1}\cr
\displaystyle{\xi}},\quad \left({d z^{*}d z}\right)^{2}=|\zeta|^{2}
d\zeta^{*}d\zeta d\xi^{*}d\xi\quad,\label{oschdash}
\end{equation}
which is related to the first solution (\ref{osch}) by an $SU(2)$
transformation
\begin{equation}
\pmatrix{\displaystyle{z_{1}}\cr
\displaystyle{z_{2}}}\mapsto\pmatrix{\displaystyle{0}&\displaystyle{i}\cr
\displaystyle{i}&\displaystyle{0}}
\pmatrix{\displaystyle{z_{1}}\cr
\displaystyle{z_{2}}}\ .\label{chngepatch}
\end{equation}
Upon this parameterization, we find another type of coherent states:
\begin{equation}
\left\vert{\xi}\right\rangle\rangle\equiv
{1\over{\left({1+|\xi|^{2}}\right)^{J}}}
\exp\left({\xi a_{1}a^{\dagger}_{2}}\right)
\left\vert{(2J,0)}\right\rangle,
\quad\xi\in{\bf C} \label{oscedash}
\end{equation}
which of course satisfy the resolution of unity
\begin{equation}
{\bf I}_{J}={{2J+1}\over{\pi}}\int{{d\xi^{*}d\xi}
\over{\left({1+|\xi|^{2}}\right)^{2}}}
\left\vert{\xi}\right\rangle\rangle
\left\langle{\langle\xi}\right\vert
\quad.\label{oscidash}
\end{equation}
The transformation from (\ref{osci}) to (\ref{oscidash}) is achieved
by
$\xi\mapsto 1/ \xi$, which is nothing but the $SU(2)$ action on ${\bf
C}P^{1}$ corresponding to the transformation (\ref{chngepatch}). It
is
now easy to see that the two critical points, $\xi=0,\ \infty$, found
in
section 2 are originated from the singular point, $(z_{1},z_{2})=0$,
of
two parameterizations (\ref{osch}) and (\ref{oschdash}).

If in (\ref{oscf}) the $\lambda$-integration is kept intact, we find
another expression of the trace formula:
\begin{eqnarray}
{\rm Tr} e^{ -iJ_{3}T}& = &
\int{{\left({dz^{*} dz}\right)^{2}}\over{\pi^{2}}}
\left\langle{\bf z}\right\vert
\int_{0}^{2\pi}{{d\lambda}\over{2\pi}}
\exp\left\{{-iT{\bf a}^{\dagger}({\sigma_{3}}/2){\bf a}
+i\lambda\left({{\bf a}^{\dagger}
{\bf a}-2J}\right)}\right\}
\left\vert{\bf z}\right\rangle\nonumber\\
& = &\lim\limits_{\varepsilon\rightarrow +0}
\int_{0}^{2\pi}{{d\lambda}\over{2\pi}}e^{-2iJ\lambda}
\int{{\left({dz^{*} dz}\right)^{2}}\over{\pi^{2}}}
\left\langle{\bf z}\right\vert
\exp\left\{{-iT{\bf a}^{\dagger}({\sigma_{3}}/2){\bf a}
+(i\lambda-\varepsilon){\bf a}^{\dagger}
{\bf a}}\right\}
\left\vert{\bf z}\right\rangle\label{trosca}\\
& = &\lim\limits_{\varepsilon\rightarrow +0}
\int_{0}^{2\pi}{{d\lambda}\over{2\pi}}e^{-2iJ\lambda}
{{1}\over{\left({1-e^{i\lambda-\varepsilon-iT/2}}\right)
\left({1-e^{i\lambda-\varepsilon+iT/2}}\right)}}\quad .\nonumber
\end{eqnarray}
By putting $w=e^{-i\lambda}$, the desired form is found;
\begin{eqnarray}
{\rm Tr} e^{ -iJ_{3}T}& = &\lim\limits_{\varepsilon\rightarrow +0}
\oint\limits_{|w|=1}{{dw}\over{2\pi i}}w^{2J+1}
{1\over{\left({w-e^{-\varepsilon-iT/2}}\right)
\left({w-e^{-\varepsilon+iT/2}}\right)}}\nonumber\\
& = &{{e^{-i(J+1/2)T}}\over{e^{-iT/2}-e^{+iT/2}}}
+{{e^{+i(J+1/2)T}}\over{e^{+iT/2}-e^{-iT/2}}}\label{troscc}\\
& = &{{e^{-iJT}}\over{1-e^{+iT}}}
+{{e^{+iJT}}\over{1-e^{-iT}}}\ .\nonumber
\end{eqnarray}
This result should be interpreted as follows: by deferring the
$\lambda$-integration the Hamiltonian (see (\ref{trosca})) is
regarded as a bilinear form of ${\bf a}^{\dagger}$ and ${\bf a}$.
Thus under the canonical coherent state representation the path
integral
can be performed to yield a product of determinants as a function of
$\lambda$. The remaining integration with respect to $\lambda$ can be
expressed as the sum of the contributions from the singular points
emerging from the reduction from ${\bf C}^{2}$ to ${\bf C}P^{1}$.

\subsection{$SU(1,1)$ case}
Introduce a set of operators
\begin{equation}
{\hat K}_{3}\equiv{1\over2}\left(a_{1}^{\dagger}a_{1}
+a_{2}^{\dagger}a_{2}+1\right)\ ,\quad
{\hat K}_{+}\equiv a_{1}^{\dagger}a_{2}^{\dagger}\ ,\quad
{\hat K}_{-}\equiv -a_{1}a_{2}\label{oscdska}
\end{equation}
and
\begin{equation}
{\hat C}\equiv a_{2}^{\dagger}a_{2}-a_{1}^{\dagger}a_{1}\ .
\label{oscdskb}
\end{equation}
The constraint
\begin{equation}
{\hat C}-(2K-1)=0\ \left(K=1/2,\ 1,\ 3/2,\ \cdots\right)
\label{oscdskc}
\end{equation}
to the Fock space gives us the $SU(1,1)$ ``{\it spin}"-$K$
representation:
\begin{eqnarray}
\left\vert{(M,2K-1+M)}\right\rangle
& = &{1\over{\sqrt{(2K-1+M)!M!}}}
\left({a^{\dagger}_{1}}\right)^{M}
\left({a^{\dagger}_{2}}\right)^{2K-1+M}
\left\vert{(0,0)}\right\rangle\nonumber\\
& \equiv & \left\vert{K,M}\right\rangle
\quad\left(M=0,\ 1,\ 2,\ \cdots\right)\ .
\label{oscdskd}
\end{eqnarray}
Therefore the $SU(1,1)$ coherent state is given as
\begin{equation}
\left\vert{\xi}\right\rangle\equiv\left({1-|\xi|^{2}}\right)^{K}
\exp\left({\xi
a^{\dagger}_{1}a_{2}^{\dagger}}\right)\left\vert{(0,2K-1)}
\right\rangle ,\quad\xi\in D_{(1,1)}\ .
\label{oscdske}
\end{equation}
As in the case of $su(2)$, the projection operator
\begin{equation}
{\bf I}_{K}\equiv\int_{0}^{2\pi}{{d\lambda}\over{2\pi}}
\exp\left({i\lambda\left\{{a_{2}^{\dagger}a_{2}-a_{1}^{\dagger}a_{1}-(2K-1)}
\right\}}\right) \label{oscdskf}
\end{equation}
is expressed in terms of the canonical coherent states as
\begin{eqnarray}
{\bf I}_K& = & \int_0^{2\pi } {{{d\lambda } \over {2\pi }}
e^{-i\lambda \left({2K-1} \right)}
\sum\limits_{m,n,Q,Q'=0}^\infty
{{{e^{i\lambda Q}} \over {m!n!\left( {m+Q} \right)!
\left( {n+Q'} \right)!}}}}\nonumber\\
\mbox{}&\times& \int {{{\left( {dz^*dz} \right)^2} \over {\pi ^2}}
e^{-{\bf z}^{\dagger}
{\bf z}}z_1^mz_2^{m+Q}{z_1^*}^n{z_2^*}^{n+Q'}{a_1^{\dagger}} ^m
{a_2^{\dagger}} ^{m+Q}\left| {\left( {0,0}
\right)} \right\rangle \left\langle {\left( {0,0} \right)}
\right|a_1^na_2^{n+Q'}}\ .\label{oscdskfa}
\end{eqnarray}
A change of variables,
\begin{equation}
\left( {{z_1},\ {z_2}} \right)=\zeta \left({
{\xi \over {\sqrt {1-\left| \xi  \right|^2}}},\ 1} \right)
,\quad \left({d z^{*}d z}\right)^{2}=|\zeta|^{2}
d\zeta^{*}d\zeta {{d\xi^{*}d\xi}\over{\left( {1-\left| \xi
\right|^2}
\right)^2}}\quad,\label{oscdskg}
\end{equation}
and the integration with respect to $\zeta$ leads us to
\begin{eqnarray}
{\bf I}_K& = &\sum\limits_{m,n=0}^\infty  {{{\left( {2K-1}
\right)!\left( {m+n+2K} \right)!} \over {m!n!\left( {m+2K-1}
\right)!\left( {n+2K-1} \right)!}}}
\left( {a_1^\dagger a_2^\dagger } \right)^m\left| {\left( {0,2K-1}
\right)} \right\rangle \left\langle {\left( {0,2K-1} \right)}
\right|\left( {a_1a_2} \right)^n\nonumber\\
\mbox{}&\times& \int\limits_{D_{\left( {1,1}
\right)}} {{{d\xi ^*d\xi } \over {\pi\left( {1-\left| \xi  \right|^2}
\right)^2}}\left( {1-\left| \xi  \right|^2} \right)^{2K+1+(m+n)/2}\xi
^m{\xi ^*}^n}\quad.\label{oscdskga}
\end{eqnarray}
Noting that the angular part of the integration in (\ref{oscdskga})
is proportional to $\delta_{m,n}$ and the identity of the
beta-function,
\begin{eqnarray}
& & B\left( {p,q} \right)B\left( {p+q,r} \right)
=B\left( {q,r} \right)B\left( {q+r,p} \right)\ ,\nonumber\\
\mbox{}& & p=2K+m,\ q=m+1,\ r=2K-1\quad,
\label{betafunct}
\end{eqnarray}
we obtain
\begin{eqnarray}
{\bf I}_{K}
& = & \sum\limits_{m,n=0}^\infty  {{{2K-1} \over
{\pi}}}\int\limits_{D_{\left( {1,1} \right)}} {{{d\xi ^*d\xi } \over
{\left( {1-\left| \xi  \right|^2} \right)^2}}
\left( {1-\left| \xi \right|^2} \right)^{2K}}\nonumber\\
\mbox{}& &\mathop{\hphantom{\sum\limits_{m,n=0}^\infty
{{1 \over{m!n!}}}\int\limits_{D_{\left( {1,1} \right)}}}}
\times{1\over{m!}}
\left( {\xi a_1^{\dagger} a_2^{\dagger} }
\right)^m\left| {\left( {0,2K-1} \right)}
\right\rangle \left\langle
{\left( {0,2K-1} \right)} \right|
{1\over{n!}}
\left( {{\xi ^*}a_1a_2} \right)^n\quad.
\label{oscdskgb}
\end{eqnarray}
In view of (\ref{oscdske}), ${\bf I}_{K}$ is
nothing but the resolution of unity in the ``{\it spin}"-$K$
representation
of $SU(1,1)$
\begin{equation}
{\bf I}_{K} ={{2K-1}\over{\pi}}\int{{d\xi^{*}d\xi}
\over{\left({1-|\xi|^{2}}\right)^{2}}}
\left\vert{\xi}\right\rangle\left\langle{\xi}\right\vert\ .
\label{oscdskh}
\end{equation}
We cannot expect, however, another type of coherent states in this
case;
because we are dealing only with one connected component of $SU(1,1)$
and there is no
$SU(1,1)$ transformation similar to (\ref{chngepatch}) leaving the
constraint (\ref{oscdskc}) invariant: for example we may use
\begin{equation}
\left( {{z_1},\ {z_2}} \right)=\zeta \left({1,\
{\xi \over {\sqrt {1-\left| \xi  \right|^2}}}} \right)
,\quad \left({d z^{*}d z}\right)^{2}=|\zeta|^{2}
d\zeta^{*}d\zeta {{d\xi^{*}d\xi}\over{\left( {1-\left| \xi
\right|^2}
\right)^2}}\quad,\label{oscdskgdash}
\end{equation}
instead of (\ref{oscdskg}). However it does provide neither $SU(1,1)$
invariant measure nor coherent states in terms of $\xi$ after
$\zeta$-integration. Therefore we have only one singularity under the
reduction ${\bf C}^{2}\rightarrow D_{(1,1)}$ in this case. Following
the same idea developed in
$SU(2)$ case, we obtain a new expression for the trace formula:
\begin{eqnarray}
& &{\rm Tr} e^{ -iK_{3}T}\nonumber\\
& = &\lim\limits_{\varepsilon\rightarrow +0}
\int_{0}^{2\pi}{{d\lambda}\over{2\pi}}
\int{{\left({dz^{*} dz}\right)^{2}}\over{\pi^{2}}}
\left\langle{\bf z}\right\vert
\exp\left\{{-i({\bf a}^{\dagger}{\bf a}+1)T/2
+i\lambda\left({-{\bf a}^{\dagger}\sigma_{3}
{\bf a}-(2K-1)}\right)
-\varepsilon{\bf a}^{\dagger}{\bf a}}\right\}
\left\vert{\bf z}\right\rangle\nonumber\\
& = &e^{-iT/2}
\lim\limits_{\varepsilon\rightarrow +0}
\int_{0}^{2\pi}{{d\lambda}\over{2\pi}}e^{-i(2K-1)\lambda}
{{1}\over{\left({1-e^{i\lambda-\varepsilon-iT/2}}\right)
\left({1-e^{-i\lambda-\varepsilon-iT/2}}\right)}}\quad
\label{troscaba}\\
& = &e^{-iT/2}\lim\limits_{\varepsilon\rightarrow +0}
\oint\limits_{|w|=1}{{dw}\over{2\pi i}}w^{2K-1}
{1\over{\left({1-we^{-\varepsilon-iT/2}}\right)
\left({w-e^{-\varepsilon-iT/2}}\right)}}\nonumber\\
& = &{{e^{-iKT}}\over{1-e^{-iT}}}\ .\nonumber
\end{eqnarray}


\begin{figure}
\caption{Contours of $z$-integration for $SU(2)$
character formula. The original contour $C$ starts from the origin
and
ends at a point in finite distance. This is the consequence of
compactness of $SU(2)$.}
\label{su2fig}
\end{figure}

\begin{figure}
\caption{Contours of $z$-integration for $SU(1,1)$ trace
formula. The original contour $C$ starts from the origin and never
returns to a point in finite distance. This is the consequence of
non-compactness of $SU(1,1)$.}
\label{su1-1fig}
\end{figure}

\end{document}